\documentclass[preprint,showpacs,preprintnumbers,amsmath,amssymb]{revtex4} %
\usepackage{graphicx}% Include figure files
\usepackage{dcolumn}% Align table columns on decimal point
\usepackage{bm}% bold math

\begin{document}
\title{Multiple scattering formalism for correlated systems: A KKR+DMFT approach}
\author{J. Min\'ar$^1$, L.Chioncel$^2$, A. Perlov$^1$, H. Ebert$^1$,
M.I. Katsnelson$^3$, and A.I. Lichtenstein$^4$}  
\affiliation{$^1$Dep. Chemie und Biochemie, Physikalische Chemie, Universit\"at M\"unchen,
  Butenandtstr. 5-13, 
 D-81377 M\"unchen, Germany \\
$^2$ Institute for Theoretical Physics and Computational Physics
Graz University of Technology, A-8010 Graz, Austria\\
$^3$University of Nijmegen, NL-6525 ED Nijmegen, The Netherlands \\
$^4$Institut f\"ur Theoretische Physik, Universit\"at Hamburg, 20355 Hamburg, Germany}
\date{\today}
\begin{abstract}
We present a charge and self-energy self-consistent computational
scheme for correlated systems based on the Korringa-Kohn-Rostoker
(KKR) multiple scattering theory with the many-body effects described by
the means of dynamical mean field theory (DMFT). The corresponding
local multi-orbital and energy dependent self-energy is included into the set of
radial differential equations for the single-site wave functions.
The KKR Green's function is written in terms of the multiple
scattering path operator, the later one being evaluated using the
single-site solution for the $t$-matrix that in turn is determined by
the wave functions. An appealing feature of this approach is that it allows to
consider local quantum and disorder fluctuations on the same
footing. Within the Coherent Potential Approximation (CPA) the
correlated atoms are placed into a combined effective medium
determined by the dynamical mean field theory (DMFT)
 self-consistency condition. Results of corresponding calculations
for pure Fe, Ni and Fe$_{x}$Ni$_{1-x}$  alloys are presented.
\end{abstract}

\maketitle

\section{Introduction}
One of the first band structure methods formulated in terms of
Green's functions is the KKR method of Koringa, Kohn and Rostoker
\cite{Kor47,KR54}. Although it is not counted among the fastest band
structure methods, it is usually regarded as a very accurate
technique. The advantage of the KKR method lies in the transparent
multiple scattering formalism which allows to express the Green's
function in terms of single-site scattering and geometrical or structural
quantities. A second outstanding feature of the KKR method is the
Dyson equation relating the Green's function of a perturbed system
with the Green's function of the corresponding unperturbed reference
system. Because of this property, the KKR Green's function method 
allows to deal with substitutional disorder including both diluted
impurities and concentrated alloys  in the framework of the
Coherent Potential Approximation (CPA) \cite{Fau82}. Within this
approach (KKR-CPA) the propagation of an electron in an alloy is regarded as
a succession of elementary scattering processes due to random
atomic scatterers, with an average taken over all configurations
of the atoms. This problem can be solved assuming that a given
scattering center is embedded in an effective medium whose choice
is open and can be made in a self-consistent way. The physical condition
corresponding to the CPA is simply that a single scatterer
embedded in the effective CPA medium should produce no further
scattering on the average. A similar philosophy is applied also
when dealing with many-body problems for crystals in the framework of
the so called dynamical mean field theory (DMFT, for review see Ref.
\onlinecite{GKKR96}).
Thus it seems to be rather natural to combine the DMFT and  KKR
methods to arrive at a very reliable and flexible band structure
scheme that include correlation effects beyond the standard local
density (LDA) or generalized (GGA) approximations. In fact the
combination of the KKR-CPA for disordered alloys and the DMFT scheme
is based on the same arguments as used by Drchal et al.\cite{DJK99}
when combining the TB-LMTO Green's function method for alloys
\cite{TDK+97} with the DMFT. In contrast to their approach, however,
the formalism presented below allows to incorporate correlation
effects via a corresponding self-energy when calculating the
electronic single-site wave functions.

First attempts to achieve a self-consistent description of local
correlation effects in crystals have been made already many years
ago. 
In the third paper of a famous series Hubbard \cite{Hub64} has
introduced an alloy analogy and by an appropriate decoupling
scheme for the Green's function a set of equations has been derived
that represent a self-consistent formulation equivalent 
to the CPA approximation.
In contrast to the DMFT the ``Hubbard III''
approximation considers quantum on-site fluctuations as
static ones which leads to some shortcomings such as violation of
some Fermi liquid properties, missing of the so called Kondo peak
near the metal-insulator transition \cite{GKKR96}. Keeping
in mind the above conceptual analogies it is our purpose to
present here a combined Local Density Approximation and Dynamical
Mean Field Theory (LDA+DMFT) electronic structure technique,
including the case of disordered solids, in the framework of the
KKR method. The many-body correlation effects are treated by means
of the DMFT, while the disorder is described in the framework
of the CPA. Taking into account the local nature of the DMFT
approximation the self-energy is represented by a local complex
energy dependent quantity (which is a matrix in orbital indices)
viewed as a contribution to the electronic potential. We note that
for a general non-local energy dependent potential multiple
scattering theory offers a solution known as the optical potential
\cite{Tay83}. However, the nonlocal self-energy is far too
complicated to be used in a realistic computation.

Very recently a combined LDA+DMFT computational scheme was proposed in
which the so called Exact Muffin-tin band structure method was used. 
In the EMTO approach \cite{AJK94,AS00,VSJ+00} the one-electron effective
potential is represented by the optimized overlapping muffin-tin
potential which is considered as the best possible spherical
approximation to the full-one electron potential. In essence the
one-electron Green's function is evaluated on a complex contour
similarly to the screened KKR technique, from which it was
derived. In the iteration procedure the LDA+DMFT Green's function is
used to calculate the charge and spin densities. Finally, for the
charge self-consistent calculation one constructs the new LDA
effective potential from the spin and charge densities
\cite{CVA+03}, using the Poisson equation in the spherical cell
approximation \cite{Vit01}.

In contrast to the EMTO implementation \cite{CVA+03}, the present
work follows a natural development in which the self-energy is
added directly to the coupled radial differential equations which
determine the electronic wave function within a potential well and
this way the single-site $t$-matrix. Because this way also the
scattering path operator of multiple scattering theory used to set up
the electronic Green's function is determined
unambiguously, no further approximations
are needed to achieve charge self-consistency.

The paper is organized as follows. Section \ref{forprob} presents
a general formulation of the problem. 
Section \ref{MSGF} provides an extension of the
derivation of the multiple scattering Green's function to include
the self-energy, and in particular provides the information on the
many-body solver. Section \ref{AIM} 
describes the many-body solver used in our calculation, that is based
on a modified fluctuating exchange interaction approximation. The
combined self-consistency cycle is presented in section
\ref{slfcns}. Finally, results and discussions are presented in
section \ref{RESULTS}.

\section{Formulation of the problem}\label{forprob}
The DMFT method has already been implemented within several band
structure methods based on a wave function formalism:
 first in the linear muffin-tin orbital method in atomic sphere
 approximation (ASA-LMTO) \cite{APK+97,LK98,KL99} and then in
full-potential LMTO \cite{SKA01,SK04}, as well as in a screened KKR or
exact muffin-tin orbitals approach (EMTO)\cite{CVA+03}. The emerged
LDA+DMFT method can be used for calculating the electronic
structure for a large variety of systems with different strength of
the electronic correlations (for a review, see Refs.
\onlinecite{LKK02,HNK+02}). To underline the importance
of complete LDA+DMFT self-consistency we mention that the first
successful attempt to combine the DMFT with LDA charge
self-consistency gave an important insight into a long-standing
problem of phase diagram and localization in f-electron systems
\cite{SKA01,SK04} and has been used also to describe correlation
effects in half-metallic ferromagnetic materials like NiMnSb
\cite{CKG+03}.
As an alternative to the above mentioned band structure methods,
accurate self-consistent methods for solving the 
local Kohn-Sham equations based on LDA in terms of Green's functions
have been 
developed within the multiple scattering theory (KKR-method)
\cite{Gyo72,Ric84,FS80,Wei90}. For that reason the KKR-method can be
combined, as it will be shown below, in a natural way with
the LDA+DMFT approach. A further appealing feature of this scheme is
that the CPA alloy theory can also be incorporated very easily.

In order to account within LDA-band structure calculations for
correlations an improved hybrid Hamiltonian
was proposed by Anisimov et al.\cite{AZA91,AAL97}. In its most general
form such a Hamiltonian is written as 

\begin {equation}
\label{Ham}
H=H_{LDA}+H_U-H_{DC}\;,
\end {equation}

where $H_{LDA}$ stands for the ordinary LDA Hamiltonian, $H_U$
describes the
effective electron-electron interaction and the one-particle
Hamiltonian $H_{DC}$ serves to eliminate double counting of the
interactions already accounted for by $H_{LDA}$.

Using second  quantization a rather general expression for $H_U$ is
given by:

\begin {equation}
\label{HamU}
H_U=\frac{1}{2}\sum_{n,ijkl}U^{n}_{ijkl}\hat C^\dagger_{ni}\hat
C^\dagger_{nj}\hat C_{nk}\hat C_{nl}\;,
\end {equation}

where $n$ runs over all the sites of the crystal $\vec R_n$ and the creation
($\hat C^\dagger$) and annihilation ($\hat C$) operators are defined with
respect to some subset of localized orbitals $\phi_i(\vec r- \vec
R_n)$. The constants $U^n_{ijkl}$ are matrix elements of the screened Coulomb
interaction $v(\vec r - \vec r\,')$:
\begin {equation}
\label{Uijkl}
U^n_{ijkl}=\int 
\phi^\dagger_i(\vec r- \vec R_n)\phi^\dagger_j(\vec r\,'- \vec R_n)
v(\vec r - \vec r\,')
\phi_k(\vec r\,'- \vec R_n)\phi_l(\vec r- \vec R_n)
d\vec rd\vec r\,' \;.
\end {equation}
The resulting many-particle Hamiltonian can not be diagonalized exactly, thus
various methods were developed in the past to find an approximate
solution \cite{GKKR96}. Among 
them one of the most promising approaches is to solve Eq.~(\ref{Ham}) within
dynamical mean field theory, a method developed  originally to deal with the
Hubbard model. 

The main idea of DMFT is to map
a periodic many-body problem onto  an effective 
single-impurity problem that has to be solved self-consistently. 
For this purpose one describes the electronic properties
of the system in terms of the one particle Green's function $\hat G(E)$,
being the solution of the equation: 

\begin {equation}
\label{dyson}
(E-\hat H-\hat \Sigma(E))\hat G=\hat 1 \;,
\end {equation}
where $E$ is the complex energy and the
effective self-energy operator $\hat \Sigma$ is assumed to be
a single-site quantity for site $n$:

\begin {equation}
\label{sigma}
\hat \Sigma(E)=\sum_{ij}|\phi_{ni}\rangle\Sigma_{ij}(E)\langle\phi_{nj}|\;.
\end {equation}

Within DMFT, the self-energy matrix $\Sigma_{ij}(E)$ is a solution of
the many-body 
problem of an impurity placed in an effective medium. This medium is
described by the so called {\em bath} Green's function matrix $\mathcal G$
defined  as:
\begin {equation}
\label {bath}
\mathcal G_{ij}^{-1}(E)=G_{ij}^{-1}(E)+\Sigma_{ij}(E)\;,
\end {equation}
where $G_{ij}(E)$ is calculated as a projection of  $\hat G(E)$ onto
the impurity site:
\begin {equation}
\label{Gproj}
G_{ij}(E)=\langle\phi_{ni}|\hat G(E)|\phi_{nj}\rangle\;.
\end {equation}

As the self-energy $\Sigma_{ij}(E)$ depends on the {\em bath} Green's function
$\mathcal G_{ij}(E)$ the DMFT equations have to be solved self-consistently. 
Accordingly, from a technical point of view the problem can be split into two
parts. One is dealing with the solution of Eq.~(\ref{dyson}) and
the second one is the effective many-body problem to find the
self-energy $\Sigma_{ij}(E)$. Within the present work, the first task is solved
by the KKR band structure method, as described below in Sec.~\ref{MSGF}.
The details of solving the
many-body effective impurity problem based 
on the fluctuation exchange (FLEX) approximation \cite{BS89a} will be
presented in Sec.~\ref{AIM}.

\subsection{The KKR+DMFT formalism}
\label{MSGF} In this section we present an extension of the well
known KKR equations in order to include the local, multi-orbital
and energy dependent self-energy produced by the many-body solver (see
section \ref{AIM}). In the framework of the multiple scattering
formalism the solution of Eq.~(\ref{dyson}) is constructed in
two steps. For the first step one has to solve the so called
single-site scattering problem, to obtain the regular ($Z$)
and irregular solution ($J$) of the corresponding Schr\"odinger (or in
our case Lippmann-Schwinger) equations as well as a scattering
amplitude expressed in terms of the single-site $t$-matrix.

\subsubsection{Solution of the single-site problem}
The solution of the single-site problem can be worked out easily
in the same way as in the full-potential description
\cite{Zel87a}. This way one finally gets the single-site $t$-matrix
for the LDA+DMFT case. In terms of the wave functions the single-site
quasiparticle equation to be solved for each spin channel $\sigma$ reads 
\begin {equation}
\label{LSe} [-\nabla^2+V^{\sigma}(r)-E]\Psi(\vec r)+\int
\Sigma^{\sigma}(\vec r,\vec r\,',E)\Psi(\vec r\,')d^{3}r\,'=0.
\end {equation}
In the following we omit the spin index $\sigma$ for the moment
keeping in mind that for a spin-polarized system described in a
non-relativistic way  one has to solve Eq.~(\ref{LSe}) for each spin
channel independently. For the solution $\Psi_\nu (\vec r)$ one can
start from the ansatz: 

\begin {equation}
\label{Ansatz}
\Psi_\nu (\vec r)=\sum_{L}\Psi_{L\nu}(\vec r)\;,
\end {equation}
where the partial waves $\Psi_{L\nu}(\vec r)$ are chosen to
have the same form as the linearly independent solutions for the
spherically symmetric potential:

\begin {equation}
\label{psi} \Psi_{L\nu}(\vec r)=\Psi_{L\nu}(r)Y_{L}(\hat r) \;,
\end {equation}
with $L=(l,m_l)$ standing for the angular momentum and magnetic
quantum numbers and $Y_L(\hat r)$ are spherical harmonics.
Inserting the ansatz (\ref{Ansatz}) into the single-site equation (\ref{LSe}) and integrating over angle variables leads to the
following set of the coupled radial integro-differential equations:
%
%eeeeeeeeeeeeeeeeeeeeeeeeeeeeeeeeeeeeeeeeeeeeeeeeeeeeeeeeeeeeeeeeeeeeeee
\begin{equation}
 \left[ \frac{d^2}{dr^2} - \frac{l(l+1)}{r^2} - V(r) + E \right]
  \Psi_{L\nu}(r,E) = \sum_{L''} \:\int r'^{2}dr' \Sigma_{LL''}(E)
  \:\phi_{l}(r)\phi_{l''}(r') \Psi_{L\nu}(r',E)\; , \label{radial} 
\end{equation}
%eeeeeeeeeeeeeeeeeeeeeeeeeeeeeeeeeeeeeeeeeeeeeeeeeeeeeeeeeeeeeeeeeeeeeee
%
For a general non-diagonal self-energy a similar radial equation (\ref{radial})
shall be written for the left-hand side equation. 
If one makes a rather natural choice of the localized subset of
functions being just $\phi_{L}(\vec r)=\phi_{l}(r)Y_L(\hat r)$ (see below).
In principle these equations can be solved by summing a corresponding
Born series. In this work, however, we simplified the equations taking advantage
of the following special representation for the self-energy:

\begin {equation}
\label{sigdelta}
\int d^3r^\prime \Sigma(\vec r,\vec r\;^\prime,E)\psi_L(\vec r\;^\prime,E)=
\sum_L\int d^3r^\prime\Sigma_{L'L}(E)\phi_{L'}^{\dagger}(\vec
r)\phi_{L}(\vec r\;') 
\psi_L(\vec r\;^\prime,E)
\approx \sum_L\Sigma_{L'L}(E)\psi_L(\vec r,E) \;.
\end {equation}

This way the Eq.~(\ref{radial}) becomes a pure differential equation:
\begin{eqnarray}
\label{RDEQUP}
\left[ \frac{d^2}{dr^2} - \frac{l(l+1)}{r^2} - V(r) + E \right]
  \Psi_{L\nu}(r,E) = \sum_{L'} \: \Sigma_{LL'}(E) \: \Psi_{L'\nu}(r,E)\; .
\end{eqnarray}

After having solved the set of coupled equations for the wave
functions one gets the corresponding single-site $t$-matrix by
introducing the auxiliary matrices $a$ and $b$ \cite{Fau82}:
\begin {eqnarray}
a_{L\nu}(E)=-ipr^2[h^{-}_{L}(pr),\Psi_{L}^\nu(r)]_r\\ \nonumber
b_{L\nu}(E)=-ipr^2[h^{+}_{L}(pr),\Psi_{L}^\nu(r)]_r\;.
\end {eqnarray}
Here $p=\sqrt{E}$ is the momentum, $h^{\pm}_{L}(pr)$ are Hankel
functions of the first and second kind and $[\ldots]_r$ denotes
the Wronskian. Evaluating the Wronskians at Wigner-Seitz radii
$r_{WS}$ one finally has \cite{Fau82,EG88}:
\begin {equation}
\label{tmat}
t(E)=\frac{i}{2p}(a(E)-b(E))b^{-1}(E)\;.
\end {equation}
The regular wave functions $Z$ used to set up the electronic Green's
function within the KKR-formalism \cite{FS80}
are obtained by a superposition of the wave
functions $\Psi_\nu$ according to the boundary conditions at
$r=r_{WS}$:
\begin {equation}
\label{zfunb}
Z_L(\vec r,E)= \sum_\nu C_L^\nu \Psi_\nu (\vec r) \stackrel{r=r_{WS}}\longrightarrow
\sum_{L\,'}j_{L\,'}(\vec r,E)t(E)^{-1}_{L,L\,'}-iph^{+}_L(\vec r,E)\;,
\end {equation}
 The irregular solutions $J_L$ needed in addition are fixed by the
 boundary condition 
\begin {equation}
\label{jfunb}
J_L(\vec r,E) \stackrel{r=r_{WS}}\longrightarrow
j_{L}(\vec r,E)
\end {equation}
and are obtained just by inward integration with the functions $j _L$
being the spherical Bessel functions.

\subsubsection{The multiple scattering Green's function}

Having constructed a set of regular ($Z$) and irregular ($J$)
solutions of the single-site problem together with the $t$-matrix the
corresponding expression for the Green's function reads \cite{FS80}:

\begin{eqnarray}
\label{KKRGF}
G(\vec r_n + \vec R_n,\vec r^{\;\prime}_{m} + \vec R_{m}, E)
&=&
\sum_{L,L'}Z_{L}(\vec
r_n,E)\tau^{nm}_{L,L'}(E)Z^{\times}_{L'}(\vec r^{\;\prime}_{m},E)\;\nonumber\\
&-&\delta_{nm}\sum_L \{ Z_{L}(\vec r_n,E)J^{\times}_{L}(\vec r^{\;\prime}_n,E)
\Theta(r^{\;\prime}_{n}-r_n)   \nonumber \\
&+&J_{L}(\vec r_n,E)Z^{\times}_{L}(\vec r^{\;\prime}_n,E) %
\Theta(r_n-r^{\;\prime}_n) \}\;.
\end{eqnarray}
Here the superscript $\times$ is used to distinguish between the left
and right hand solutions to Eq.~(\ref{LSe}); i.e. for example $|Z>$
and $<Z^{\times}|$ are solutions to the adjoint equations
\cite{Tam92}:
\begin{eqnarray}
(\hat H +\hat \Sigma -E )|Z\rangle &=& 0 \\
\langle Z^\times|(\hat H+\hat \Sigma-E)&=& 0 \;.
\end{eqnarray}

The central quantity in Eq.~(\ref{KKRGF}) 
is the scattering path operator $\tau$ which
for the case of a periodic crystal can be obtained from the Brillouin zone
(BZ) integration:

%
%%%%%%%%%%%%%%%%%%%%%%%%%%%%%%%%%%%%%%%%%%%%%%%%%%%%%%%%%%%
\begin{equation}\label{TAUBZ}
  \tau^{nm}_{LL'}(E)=\frac{1}{V_{\rm BZ}}
         \int_{V_{\rm BZ}}{\rm d}^3 k
   \left[ t^{-1}(E)-G(\vec k,E)
   \right]^{-1}_{L'L} e^{i\vec k\vec R_{nm}}
       \ ,
\end{equation}
%%%%%%%%%%%%%%%%%%%%%%%%%%%%%%%%%%%%%%%%%%%%%%%%%%%%%%%%%%%
%
where $V_{\rm BZ}$ is the volume of the first Brillouin--zone and
$\vec R_{nm}=\vec R_{n}-\vec R_{m}$ with $\vec{R}_{n(m)}$
denoting the lattice vector specifying the position of the unit cell
$n(m)$ and the matrix
$t^{-1}(E)-G(\vec k,E)$ occurring in the integral
is known as the KKR matrix.  The matrix $G(\vec k,E)$ is the Fourier
transform of the real space KKR structure constant matrix that depends
only on the relative positions of scatterers.

Given the local nature of the many-body solver used within the
DMFT approach, the KKR Green's function (\ref{KKRGF}) has to be
projected accordingly to the matrix $G^{nm}_{LL\;\prime}$ 
(see Eq.~(\ref{Gproj})). The
projection is performed through the following integration:

\begin{eqnarray}
\label{PROJGF}
G^{nm}_{L,L'}(E)=\sum_{L_1, L_2} \left( \int d^3r_{1}
  \phi^{\dagger}_{L}
({\bf r_{1}}) Z_{L_1}({\bf r_1},E) \right) \tau^{nm}_{LL'}(E)
\left( \int d^3r_{2} Z_{L_2}^{\times}({\bf r_2},E) \phi_{L\;\prime}({\bf r_2}) \right) \nonumber \\
-\delta_{nm} \sum_{L_1} ( \int d^3r_2 \left( \int_0^{r_2} d^3r_1  \phi^{\dagger}_{L}
({\bf r_1}) Z_{L_1}({\bf r_1},E) \right)
 J^{\times}_{L_1} ({\bf r_2},E) \phi_{L'} ({\bf r_2}) \nonumber \\
+ \int d^3r_2 \left( \int_{r_2}^{r_{ws}} d^3r_1 \phi^{\dagger}_{L}
({\bf r_1}) J_{L_1}({\bf r_1},E)  \right) Z^{\times}_{L_1} ({\bf r_2},E) \phi_{L'}
({\bf r_2}) )\; .
\end{eqnarray}

The impurity Green's function $G^{nm}_{LL'}(E)$ (actually
$G^{\sigma nm}_{LL'}(E)$ for both spin channels) represents
the input into the solution of the effective impurity
problem presented below.
As the DMFT-approach (see next section) concentrates on the
correlation among electrons of the same angular momentum $l$ only the
$l-l$-subblock of this matrix will be used in the following. For the
transition metal systems dealt here this implies that only the
$d-d$-subblock is considered with $\phi_L(\vec r)$ being appropriate
reference wave functions with $l=2$. 

\subsection{Solution of the effective impurity problem}\label{manys}
\label{AIM}

Our approach to achieve a solution of the many-body effective impurity
problem is based on the fluctuation exchange (FLEX) approximation
\cite{BS89a} but with a different treatment of particle-hole
and particle-particle channels. The particle-particle channel is
described by a $T$-matrix approach \cite{Gal58,Kan63} giving a
renormalization of the effective interaction, the latter one being
used explicitly in the particle-hole channel
\cite{KL99,KL02}.

The symmetrization of the {\it bare U } matrix is done over
particle-hole and particle-particle channels:
\begin{eqnarray}
U^d _{m_1 m_3 m_2 m_4} &=&2U_{m_1 m_2 m_4 m_3}^i
 - U_{m_1 m_2 m_3 m_4}^i \nonumber  \\
U^m _{m_1 m_3 m_2 m_4}  &=& - U_{m_1 m_2 m_3 m_4 }^i \nonumber  \\
U^s _{m_1 m_3 m_2 m_4} &=& \frac 1 2 (
U_{m_1 m_3 m_2 m_4}^i + U_{m_1 m_3 m_4 m_2}^i)
\nonumber \\
U^t _{m_1 m_3 m_4 m_2 } &=& \frac 1 2 (
U_{m_1 m_3 m_2 m_4 }^i - U_{m_1 m_3 m_4 m_2 }^i )\;.
\nonumber
\end{eqnarray}
As indicated above, here and in the following only matrix elements
with respect to the $d$-like reference wave functions $\phi_L$ have to
be considered.
The above expressions are the matrix elements  of bare interaction
which can be obtained with the help of the pairwise operators
corresponding to different channels:
\begin{itemize}
\item particle-hole density
\begin{equation}
d_{12} =  \frac 1 {\sqrt2 } ( c_{1 \uparrow}^+ c_{2 \uparrow} +
c_{1 \downarrow} ^+ c_{2 \downarrow} )
\end{equation}
\item particle-hole magnetic
\begin{eqnarray}
m_{12}^0 &=& \frac 1 {\sqrt2 } ( c_{1 \uparrow}^+ c_{2 \uparrow} -
c_{1 \downarrow} ^+ c_{2 \downarrow} ) \nonumber  \\
m_{12}^+ &=& c_{1 \uparrow} ^+ c_{2 \downarrow} \\
m_{12}^- &=& c_{1 \downarrow} ^+ c_{2 \uparrow} \nonumber
\end{eqnarray}
\item particle-particle singlet
\begin{eqnarray}
s_{12} &=& \frac 1 {\sqrt2 } ( c_{1 \downarrow} c_{2 \uparrow} -
c_{1 \uparrow}  c_{2 \downarrow}) \nonumber \\
\overline{s}_{12} &=& \frac 1 {\sqrt2 } ( c_{1 \uparrow}^+ c_{2 \downarrow}^+ -
c_{1 \downarrow} ^+ c_{2 \uparrow}^+)
\end{eqnarray}
\item particle-particle triplet
\begin{eqnarray}
t_{12}^0 &=& \frac 1 {\sqrt2 } ( c_{1 \downarrow} c_{2 \uparrow} +
c_{1 \uparrow}  c_{2 \downarrow}) \nonumber \\
\overline{t}_{12}^{0} &=& \frac 1 {\sqrt2 } ( c_{1 \uparrow}^+ c_{2 \downarrow}^+ +
c_{1 \downarrow} ^+ c_{2 \uparrow}^+ ) \nonumber \\
t_{12}^{\pm} &=& c_{1 \uparrow, \downarrow} c_{2 \downarrow, \uparrow} \nonumber \\
\overline{t}_{12}^{\pm} &=& c_{1 \uparrow, \downarrow}^+ c_{2 \downarrow,
  \uparrow}^+ \;.
\end{eqnarray}
\end{itemize}
These operators describe the correlated movement of the electrons
and holes below and above the Fermi level and play an
important role in defining the spin-dependent effective potentials
$W_{m_1 m_2 m_3 m_4 }^{\sigma \sigma^{\;\prime}}$. The one-electron
Green's function matrix containing the many-body interaction, described by
the self-energy $\Sigma_{m m^{\;\prime} \sigma }(i \omega_n)$ is
given by the Dyson equation
\begin{equation} \label{Green}
\mathcal G_{m m^{\;\prime} \sigma}^{-1}(i \omega_n)= (i \omega_n + \mu )
\delta_{m m^{\;\prime}}
-h_{m m^{\;\prime} \sigma} - \Sigma_{m m^{\;\prime} \sigma }(i \omega_n)
\end{equation}
where $\mu$ is the chemical potential, $\omega_n = (2n+1)\pi/\beta$
are Matsubara frequencies and $\beta = 1/T$ is the inverse
temperature. The GW type of diagrams are summed up
self-consistently to produce the self-energy. For getting the
self-energy we use a two-step FLEX approximation. This means that
first 
of all the bare matrix vertex is replaced by the $T$-matrix approach
\cite{Gal58,Kan63} which will be used in the calculation of the
particle-hole channel. In the Kanamori $T$-matrix approach the sum
over the ladder graphs may be carried out with the aid of the so
called $T$-matrix which obeys the Dyson-like integral equation:
\begin{eqnarray}
< 1 3 | T^{\sigma \sigma^{\;\prime}}(i \Omega) |24 >&= &    \nonumber \\
< 1 3 | v |24 >&-& \frac 1 \beta \sum_{\omega} \sum_{5678}
< 1 3 | v |5 7 > \mathcal G_{56} ^{ \sigma} ( i \omega)
\mathcal G_{78} ^{ \sigma^{\;\prime}} (i \Omega - i \omega )
< 68 | T^{\sigma \sigma^{\;\prime}} (i \Omega )|24 > \;.\nonumber
\end{eqnarray}
The Hartree and Fock contribution are obtained replacing the bare interaction
by a {\it T-matrix}:
\begin{eqnarray}
\Sigma_{12, \sigma}^{(TH)} (i \omega) &=& \frac 1 \beta \sum_{\Omega}
\sum_{ 34 \sigma^{\prime}}
<13|T^{\sigma \sigma^{\prime}} (i \Omega) |24>
\mathcal G_{43}^{ \sigma^{\prime}}(i\Omega -i \omega )   \\
\Sigma_{12, \sigma}^{(TF)} (i \omega) &=& - \frac 1 \beta \sum_{\Omega}
\sum_{ 34 }
<14|T^{\sigma \sigma} (i \Omega) |32>
\mathcal G_{34}^{ \sigma}(i\Omega -i \omega )
 \ .
\end{eqnarray}
In the low-density limit the self-energy should be the summation over diagrams
for repulsion of two holes below $E_F$ (ladder approximation).
Going beyond the low density limit means the inclusion of
excitations of electrons from states below the Fermi level into
the unoccupied part of the $d$ band. This process renormalizes the
hole states below $E_F$ and put new poles for the Green's function.

Combining the density and the magnetic parts of the  particle-hole
channel we can write the expression for the interaction part of
the Hamiltonian \cite{KL99,KL02}:
\begin{eqnarray}
H_{U}&=&\frac 1 2 Tr( D^+ *V^{\|}*D+ m^+*V_m
^{\bot}*m^{-}+m^-*V_m^{\bot}*m^+ )\; ,
\end{eqnarray}
where $D$ is a row matrix with elements $(d, m^0)$, and $D^+$ is a column
matrix with elements $(d^+ m_0^+)$. We denote by * matrix multiplication
with respect to the pairs of orbital indices. The expression for the
effective potential is:
\begin{eqnarray}
V^{\|}(i\omega) & = & \frac 1 2\left(\begin{array}{cc}
 {V}^{dd } &
 {V}^{dm } \\
 {V}^{md } &
 {V}^{mm}
\end{array} \right) \\
(V_m^{\bot})_{1234} & = & < 13|T^{\uparrow \downarrow}|42> \;.
\end{eqnarray}
The matrix elements of the effective interaction for $z$ or longitudinal
spin-fluctuations
are:
\begin{eqnarray}
V_{dd}&=&\frac{1}{2}\sum_{\sigma} ( \sum_{ \sigma^{\;\prime}}
<13|T^{\sigma \sigma^{\;\prime}}|42> -
<13|T^{\sigma^{\;\prime}\sigma^{\;\prime}}|42> ) \nonumber \\
V_{dm } = V_{md } &=& \frac {1} {2} \sum_{ \sigma \sigma^{\;\prime}}
  \sigma ( <13|T^{\sigma \sigma}|42> - <13|T^{\sigma \sigma}|2 4>  +
<13|T^{\sigma^{\;\prime}\sigma} |42>) \nonumber  \\
V_{mm}&=&\frac{1}{2} \sum_{\sigma} ( \sum_{ \sigma^{\;\prime}}
\sigma \sigma^{\;\prime} <13|T^{\sigma \sigma^{\;\prime}}|42> -
<13|T^{\sigma^{\;\prime}\sigma^{\;\prime}}|42> ) \nonumber \;.
\end{eqnarray}
For finite temperature
the definition for the spin dependent Green's function is:
\begin{eqnarray}
\mathcal G_{12}^{ \sigma}(\tau) &=& - < T_{\tau}c_{1\sigma}(\tau)c_{2\sigma}^+ (0) > \nonumber \\
\mathcal G_{12}^{ \sigma}(i \omega_n ) &=& \int_0^\beta e^{i \omega_n \tau}
\mathcal G_{12}^{ \sigma}(\tau) d \tau \nonumber \;.
\end{eqnarray}
The corresponding expressions for the generalized 
longitudinal $ \chi^{\|}$ and transversal  $ \chi^{\bot}$
susceptibilities are:
\begin{eqnarray}
\chi^{\bot}(i \omega)&=&[1+V_m^{\bot}
\Gamma^{\uparrow \downarrow}(i \omega)]^{-1}
* \Gamma^{\uparrow \downarrow}(i \omega) \\
\chi^{\|}(i \omega) &=& [1+ V^{\|}*\chi_0^{\|}(i \omega) ]^{-1} *
\chi_0^{\|} (i \omega) \;,      
\end{eqnarray}
where $\Gamma(i \omega)$ represent the Fourier transform of the empty
loop:
\begin{equation}
\Gamma_{m_1 m_2 m_3 m_4}^{\sigma \sigma^{\;\prime}} (\tau) = -
\mathcal G_{m_2 m_3}^{ \sigma} (\tau) \mathcal G_{m_4 m_1}^{ \sigma^{\;\prime}} (-\tau)
\end{equation} 
and the matrix of the bare longitudinal susceptibility is:
\begin{equation}\label{chilo}
\chi_0^{\|}(i\omega)=\frac 1 2\left(\begin{array}{cc}
 {\Gamma}^{\uparrow \uparrow }+{\Gamma}^{\downarrow \downarrow} &
 {\Gamma}^{\uparrow \uparrow }-{\Gamma}^{\downarrow \downarrow} \\
 {\Gamma}^{\uparrow \uparrow }-{\Gamma}^{\downarrow \downarrow} &
 {\Gamma}^{\uparrow \uparrow }+{\Gamma}^{\downarrow \downarrow} 
\end{array} \right)\;.
\end{equation}
The four matrix elements of the bare longitudinal susceptibility
represent the density-density $(dd)$, density-magnetic $(dm^0)$,
magnetic-density $(m^0d)$ and magnetic-magnetic channels $(m^0m^0)$.
The matrix elements couple longitudinal magnetic fluctuations with density
magnetic fluctuations. In this case the particle hole contribution to the
self-energy is:
\begin{equation}\label{selfph}
\Sigma_{12 \sigma} ^{(ph)}(\tau) = \sum_{34 \sigma^{\;\prime}} W_{1342}^{\sigma
\sigma^{\;\prime}} (\tau) \mathcal G_{34}^{\sigma^{\;\prime}} (\tau)  \;,
\end{equation} 
with the particle-hole fluctuation potential matrix
\begin{equation}
W ^{\sigma \sigma^{\;\prime}} (i \omega) = \left( \begin{array}{cc}
 {W}_{\uparrow \uparrow } &  {W}_{\uparrow \downarrow} \\
 {W}_{\downarrow \uparrow } &  {W}_{ \downarrow \downarrow}
\end{array} \right) \;,
\end{equation}
and the spin-dependent effective potentials defined as:
\begin{eqnarray}
W_{\uparrow \uparrow } &=& \frac {1} {2} V^{\|} *
(\chi^{\|}-\chi_0^{\|})*V^{\|} \nonumber  \\
W_{\downarrow \downarrow } &=& \frac {1} {2}V^{\|}*
( \tilde \chi^{\|} - \tilde \chi_0^{\|} )*V^{\|} \nonumber \\
W_{\uparrow \downarrow } &=& \frac{1}{2} V_m^{\bot}
*(\chi^{+-}-\chi_0^{+-} )* V_m^{\bot}  \nonumber \\
W_{\downarrow \uparrow} &=& \frac{1}{2}V_m^{\bot}
*(\chi^{-+}-\chi_0^{-+} )* V_m^{\bot}
\end{eqnarray}
The definitions for $\tilde \chi^{\|}$ and $\tilde \chi_0^{\|} $ differ
from those of
$ \chi^{\|}$ and $\chi_0^{\|} $ , respectively, by the replacement
$\Gamma^{\uparrow \uparrow} \leftrightarrow
\Gamma^{\downarrow \downarrow}$
in Eq.~(\ref{chilo}).
The complete expression for the self-energy is finally given by:
\begin{equation}
\Sigma = \Sigma^{(TH)}+ \Sigma^{(TF)} + \Sigma^{(ph)}\;.
\end{equation}

The attractive feature of the present approach is that it leads to
an exact expression for the self-energy in the limit of a small
number of holes in the $d$ band. These conditions are satisfied
with high accuracy in the case of Ni. Further details and
justifications of this approach can be found in Ref.
\onlinecite{KL02}.

\subsection{Treatment of disordered alloys}
\label{CPA+DMFT} In this section we review the KKR-CPA approach
and present a simple and transparent electronic theory that
combines the treatment of disorder and correlation on the same footing. After
several decades of intense research the problem of
interacting electrons in disordered alloys still induce numerous
investigations both 
experimentally and theoretically. In the weakly disordered limit
\cite{LR85} both disorder and interaction can be treated in a
perturbative way; note that this perturbation theory is not
trivial, in particular, a non-Fermi-liquid behaviour appears. For
strong disorder Anderson localization effects eventually lead to
the breakdown of the metallic phase and a metal-to-insulator
transition takes place (for a review, see Ref. \onlinecite{BK94}).
It was realized recently that the Hubbard model can be solved
exactly in the limit of infinite space dimensionality $d=\infty$
and in this case the Mott metal-insulator transition can be
described in the framework of dynamical mean-field theory (for a
review, see Ref. \onlinecite{GKKR96}). The presence of disorder
in $d=\infty$ increases the complexity of the problem: the cavity
field varies from site to site reflecting the random environments in
which a given site is embedded \cite{DK94}.
Fortunately, for $d=\infty$ the problem can be simplified due to a
(infinitely) large number of neighbours, in this case the cavity
fields become independent of disorder and only local disorder
fluctuations survive. We will adopt this approach which is
flexible enough to allow for study numerous interesting questions
in connection with an interplay between correlations and local
disorder \cite{Sov67}. Furthermore it is supported by the arguments
given by Drchal et al. \cite{DJK99}. 
These authors pointed out that an averaged coherent
potential for disordered interacting systems can be constructed using
the so-called terminal-point approximation. Using a local mean-field
approximation to treat electron correlations, the corresponding
self-energy gets diagonal in the site representation. This allows to use
the coherent potential alloy theory (CPA) \cite{Sov67} for the
configurational averaging in the usual way.

Among the electronic structure theories, those based on the
multiple scattering formalism are the most suitable to deal with
disordered alloys within the coherent potential approximation
(CPA). CPA is considered to be the best theory among
the so-called single-site (local) alloy theories that assume
complete random disorder and ignore short-range order
\cite{Fau82}. Combining the CPA with multiple scattering theory
leads to the KKR-CPA scheme, which is applied nowadays extensively for
quantitative investigations of the electronic structure and
properties of disordered alloys \cite{Fau82,EA92}. Within the CPA
the configurationally averaged properties of a disordered alloy
are represented by a hypothetical ordered CPA-medium, which in
turn may be described by a corresponding site-diagonal ($n=m$)
scattering path operator $\tau^{\rm CPA}$. The corresponding single-site
$t$-matrix $t^{\rm CPA}$ and multiple scattering
path operator $\tau^{{\rm CPA}}$ are determined by
the so called CPA-condition:
 %
 %%%%%%%%%%%%%%%%%%%%%%%%%%%%%%%%%%%%%%%%%%%%%%%%%%%%%%%%%%%%%%
 \begin{equation}\label{CPA-EQ}
 x_{\rm A} \tau^{\rm A} +
       x_{\rm B} \tau^{{\rm B}} =
            \tau^{{\rm CPA}}.
 \end{equation}
 %%%%%%%%%%%%%%%%%%%%%%%%%%%%%%%%%%%%%%%%%%%%%%%%%%%%%%%%%%%%%%
 %
Here a binary system A$_{x}$B$_{1-x}$ composed of components A and
B with relative concentrations $x_{\rm A}$ and $x_{\rm B}$ is
considered. The above equation represents the requirement that
embedding substitutionally an atom (of type A or B) into the CPA
medium should not cause additional scattering. The scattering
properties of an A atom embedded in the CPA medium, are
represented by the site-diagonal ($n=m$) component-projected
scattering path operator 
 $\tau^{{\rm A}}$
 %
 %%%%%%%%%%%%%%%%%%%%%%%%%%%%%%%%%%%%%%%%%%%%%%%%%%%%%%%%%%%%%%
 \begin{equation}\label{PROJTAU}
 \tau^{{\rm A}} = \tau^{{\rm CPA}}
         \left[ 1 +
                   \left( t_{\rm A}^{-1} -
                          t_{\rm CPA}^{-1}
                   \right)
                  \tau^{{\rm CPA}}
         \right]^{-1}\enspace,
 \end{equation}
 %%%%%%%%%%%%%%%%%%%%%%%%%%%%%%%%%%%%%%%%%%%%%%%%%%%%%%%%%%%%%%
 %
 where $t_{\rm A}$ and $t_{{\rm CPA}}$ are the
 single-site matrices of the A component and of the CPA effective
 medium. A corresponding  equation holds also for the B component in
 the CPA medium. The coupled sets of equations 
 for $\tau^{{\rm CPA}}$ and $t^{\rm CPA}$ have to be
 solved iteratively within the CPA cycle.

It is obvious that the above scheme can straightforwardly be extended to include the
many-body correlation effects for disordered alloys. As was pointed
out in Sec.~\ref{MSGF}, within the KKR+DMFT approach the
local multi-orbital and energy dependent self-energy
($\Sigma_A(E)$ and  $\Sigma_B(E)$) is directly included in the
single-site matrices $t_A$ and  $t_B$, 
respectively. Having solved the CPA equations self-consistently, one
has to project the CPA Green's function onto the components $A$ and
$B$ by using Eqs.~(\ref{PROJGF}) and (\ref{PROJTAU}). In
Eq.~(\ref{PROJGF}) the multiple 
scattering path operator $\tau_{LL'}^\sigma(E)$ has to be replaced
by the component-projected scattering path operator
$\tau_{LL'}^{{\rm A},\sigma}$ of an A-atom in a CPA medium. The components
Green's functions $G_{i=A,B}$ are used to construct the
corresponding bath Green's functions for which the DMFT
self-consistency condition is used according to Eq.~(\ref{bath}):
\begin {equation}
\mathcal G^{-1}_{i=A,B}(E)= G^{-1}_{i=A,B}(E)+\Sigma_{i=A,B}(E)\;.
\end {equation}
The many-body solver presented in section \ref{AIM} in turn is used to
produce the component specific self-energies $\Sigma_{i=A,B}(E)$:
\begin {equation}
\Sigma_{i=A,B}(E)=\Sigma_{i=A,B}(E)[\mathcal G_{i=A,B}(E)]\;.
\end {equation}

\subsection{The self-consistency cycle}
\label{slfcns} 
\begin{figure}
\begin{center}
\includegraphics[angle=90,height=12cm]{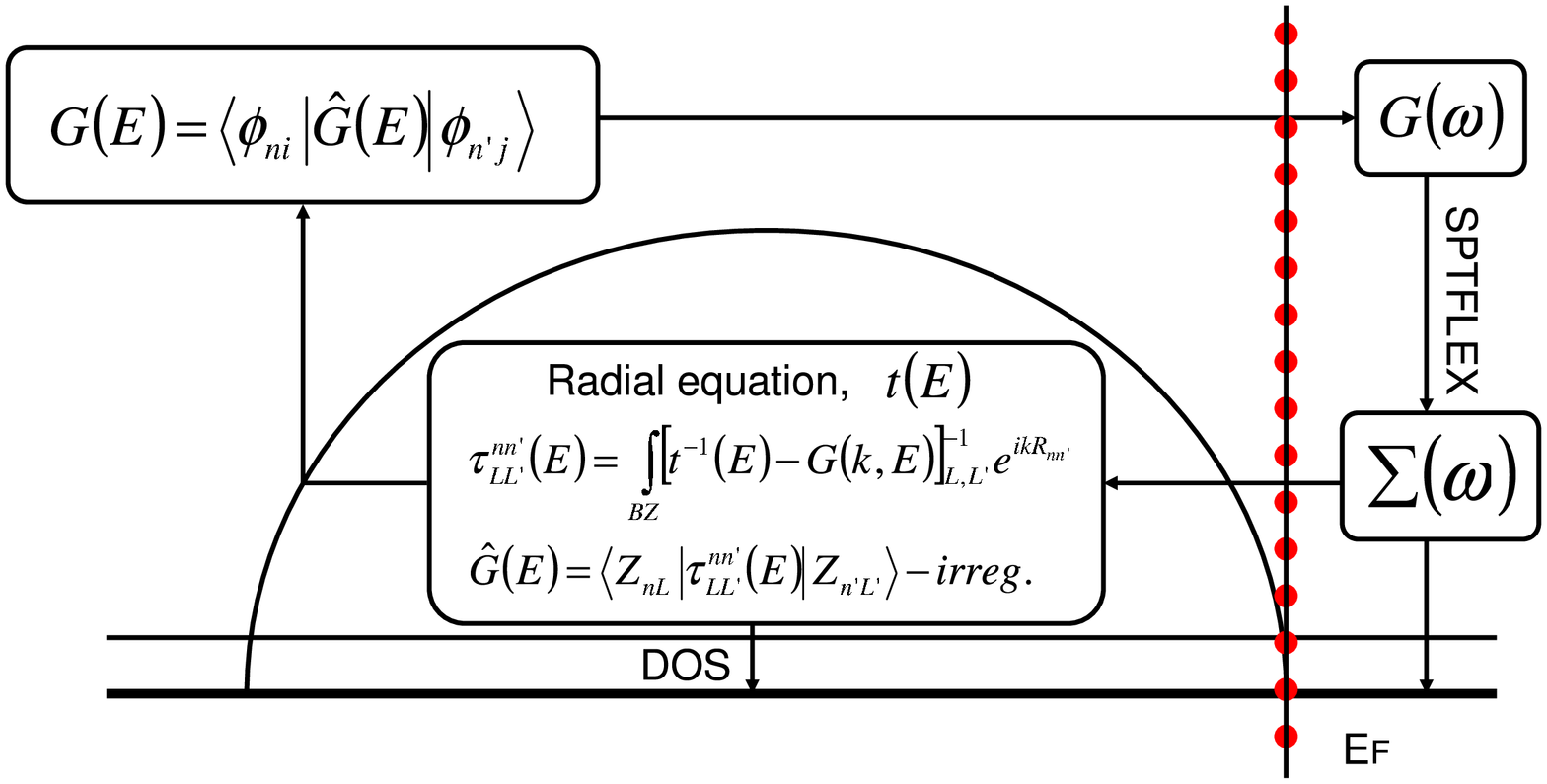}
\end{center}
\caption{The complex energy contours used within the selfconsistent LDA+DMFT
approach, combined with the KKR formalism. The solution of the radial equation
allows the evalaution of the single site scattering matrix $t(E)$
and the scattering path operator $\tau^{nn'}_{L,L'}(E)$ from which the KKR Green
function is constructed. The projection of the Green function is
perfomed according to Eq.~(\ref{PROJGF}). The impurity Green function is
then used to solve the many-body problem within the spin polarized 
$T$-matrix FLEX solver of the DMFT approach.}
\label{FIG:contour}
\end{figure}

Finally a description of the flow diagram of the
self-consistent LDA+DMFT approach is presented in Fig.
(\ref{FIG:contour}). The radial equation Eq. \ref{radial}
provides the set of regular $(Z)$ and $(J)$ irregular solutions of the
single-site problem. Together with the $t$ matrix, the scattering
path operator $\tau$ Eq.(\ref{TAUBZ}) and the KKR Green's function is
constructed Eq. \ref{KKRGF}. To solve the many-body problem the a projected
impurity  Green's function is constructed according to Eq. \ref{PROJGF}.
The LDA Green's function
$G^{nn}_{LL'}(E)$ is calculated on the complex
contour which encloses the valence band one-electron energy poles.
The Pad\'e analytical continuation is used to map the complex
local Green's function $G^{nn}_{LL'}(E)$ on the Matsubara axis which is
used when dealing with the many-body problem. In the current
implementation the perturbative SPTF (spin-polarized $T$-matrix +
FLEX) solver of the DMFT problem described above is used. In fact
any DMFT solver could be included which supplies the self-energy
$\Sigma(\omega)$ as a solution of the many-body problem. The 
Pad\'e analytical continuation is used once more to map back the
self-energy from the Matsubara axis to the complex plane, where
the new local Green's function is calculated. As was described in
the previous sections, the key role is played by the scattering
path operator $\tau^{nn}_{L,L'}(E)$, which allows us
to calculate the charge at each SCF iteration and the new potentials
that are used to generate the new LDA Green's function. In practice
it turns out that the self-energy converges faster than the charge density. 
Of course double counting corrections have to be considered
explicitly when calculating the total energy (not done here). 
Concerning the self-energy used here the double counting
corrections are included when solving the many-body problem (see
Ref.~\onlinecite{KL02}).

\section{Results and Discussion}\label{RESULTS}
 To demonstrate the capability of our approach we first applied it to
 the $3d$ metals Ni and 
 Fe. Although these metals are more or less adequately described in the
 framework of standard LDA, nevertheless, there are some
 features in the experimental properties which are due to correlation
 effects that are not adequately described on this basis. In addition, there are
 numerous investigations in the literature that seek for an improved
 description of correlation  effects in these systems and that can be
 compared with. 

\subsection {Numerical details}

The self-consistent LDA+DMFT calculations were carried out for the
experimental ground state crystal structures, i.e.  fcc for Ni, 
bcc for Fe and fcc for Ni rich Fe$_{x}$Ni$_{1-x}$ alloys. 
The lattice parameters were fixed at the experimental
values (Fe: 5.406 a.u., Ni: 6.658 a.u., for 
Fe$_{x}$Ni$_{1-x}$: see Ref.\onlinecite{Was90}).
The Green's function was calculated for 32 energy points 
distributed over semicircular contour. The Brillouin
zone integration has been performed on a uniform grid, taking
into account the symmetry of the system. 
As a suitable reference wave functions $\phi_L(\vec r-\vec R)$ we
 have choosen a radial solution of the Schr\"odinger equation for the
spherically symmetric LDA non-magnetic potential, that is determined
for an appropriate energy (E=0.7 Ry).
The DMFT parameters, average Coulomb
interaction $U$, exchange energy $J$, and temperature $T$ used in the
calculations are listed in Table \ref{TAB1}.

\subsection {Results for  bcc-Fe and fcc-Ni}
To demonstrate the applicability of the scheme presented above
band-structure calculations for bcc-Fe and fcc-Ni have been
performed. The results of LDA+DMFT calculation for both systems have
been already several times discussed in detail in the literature
\cite{DJK99,LKK01,CVA+03}. 

\begin{figure}
\begin{center}
  \includegraphics[height=10cm]{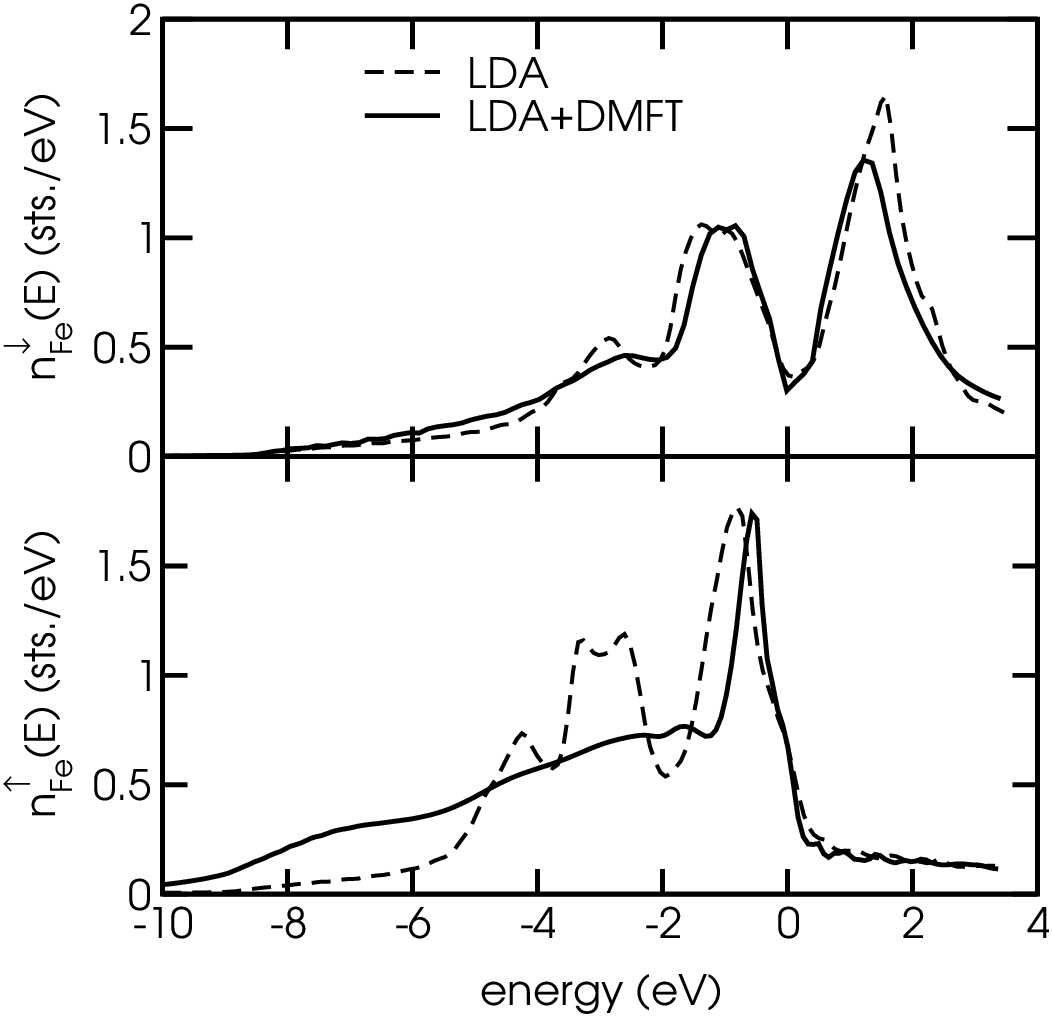}
\end{center}
\caption{Spin resolved density of states of bcc-Fe as calculated within
  LDA(dashed line)  and LDA+DMFT (full line) using the KKR-method. 
(DMFT parameters: U=2eV, J=0.9eV, T=400K) }
\label{FIG:FeDOS}
\end{figure}

The density of states curves resulting from a plain LDA and a
LDA+DMFT calculations are shown in Figs.~\ref{FIG:FeDOS} and
\ref{FIG:NiDOS} for Fe and Ni, respectively. For the LDA+DMFT
calculations we used the DMFT parameters as given in Table~\ref{TAB1}. The
density of states curves for Fe and Ni are in reasonable agreement
with corresponding previous LMTO+DMFT \cite{DJK99}, as well as EMTO+DMFT
\cite{CVA+03} calculations. The same is true also for the spin magnetic
moments (see Table \ref{TAB1}). The spin magnetic moments
are some what higher in comparison with the EMTO+DMFT
results.\cite{CVA+03}
From Figs.\ref{FIG:FeDOS} and \ref{FIG:NiDOS} one can see that in
bcc-Fe the correlation effects are much less pronounced than in
fcc-Ni. This is due to the large exchange splitting for Fe and the
bcc-structure dip in the minority density of states \cite{LKK01}. In
the case of Ni the LDA+DMFT calculations account for all expected
influences of
 the density of states in satisfying way. As can be seen from
Fig.~\ref{FIG:NiDOS}, the density of states reflects all three main
correlation effects: the $30\%$ narrowing of the occupied part of the
d-band, about $40\%$ decrease of exchange splitting and the presence
of the famous 6eV satellite compared to the LDA DOS. However, the
position of the 6eV sattelite is shifted somewhat to lower binding
energies. This shift and the large broadening of the resonance is due
to the perturbation approach of the DMFT solver of the effective
impurity problem used here.\cite{KL02}
\subsection {Results for  fcc-Fe$_{x}$Ni$_{1-x}$ disordered alloy}

\begin{figure}
\begin{center}
  \includegraphics[height=10cm]{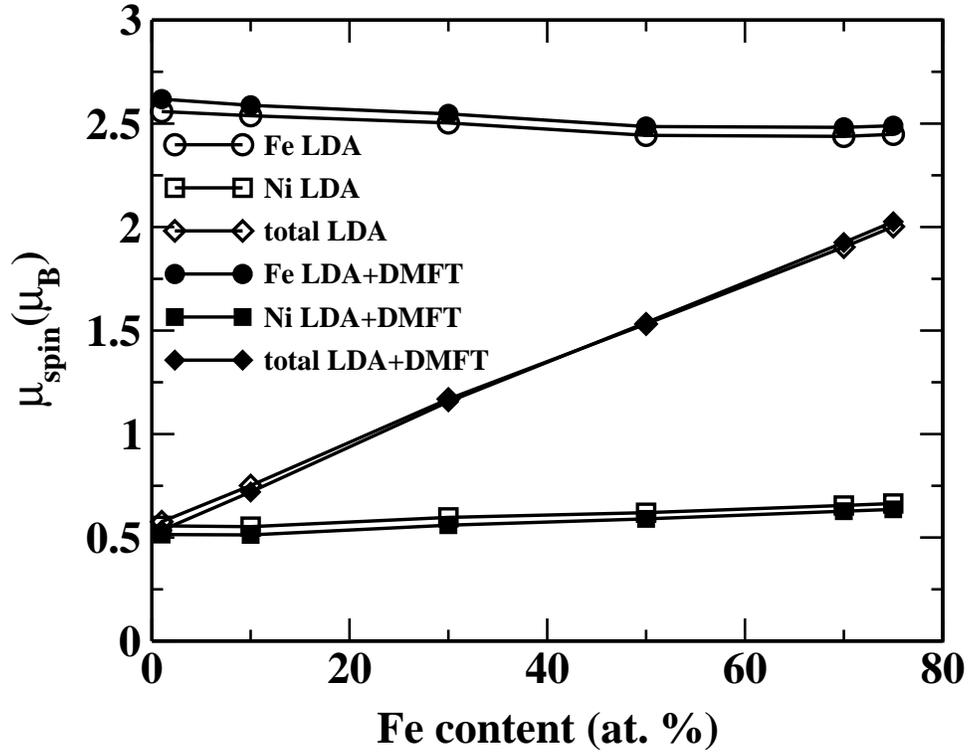}
\end{center}
\caption{Spin magnetic moments of Fe (circles) and Ni (squares) in
  Fe$_{x}$Ni$_{1-x}$ alloy calculated using plain LDA (open symbols)
  and the LDA+DMFT method (full symbols). }
\label{FIG:FeNiMAG}
\end{figure}

As mentioned above, the scheme presented here allows in a straight
forward way to deal with disordered alloys. To demonstrate how this
works we carried out a set of LDA as well as LDA+DMFT calculations
for fcc-Fe$_{x}$Ni$_{1-x}$ disordered alloy for various
concentrations. For the LDA+DMFT we used the same DMFT parameters (U,
J and T) as in the case of pure bcc-Fe and fcc-Ni (see
Table~\ref{TAB1}). In Fig.~\ref{FIG:FeNiMAG} the element resolved as well
as the total spin magnetic moments are shown. Although the difference
between LDA and LDA+DMFT moments for Fe is rather small one can see
an interesting trend. In contrast to the pure bcc-Fe case the LDA+DMFT
moments for Fe in fcc-Fe$_x$Ni$_{1-x}$  alloy are slightly larger than
corresponding LDA ones. In the case of Ni, on the other hand, 
a decrease of the magnetic moment was
obtained as in the case of pure fcc-Ni (see Table~\ref{TAB1}). Comparing
the average moments in Fig.~\ref{FIG:FeNiMAG} with experiment
\cite{Was90} one finds rather good agreement already for LDA-based spin
moment. In spite of its present limitations, the LDA+DMFT scheme does not
spoil the overall behaviour for the concentration dependence of magnetic
moments.

\begin{figure}
\begin{center}
  \includegraphics[height=12cm]{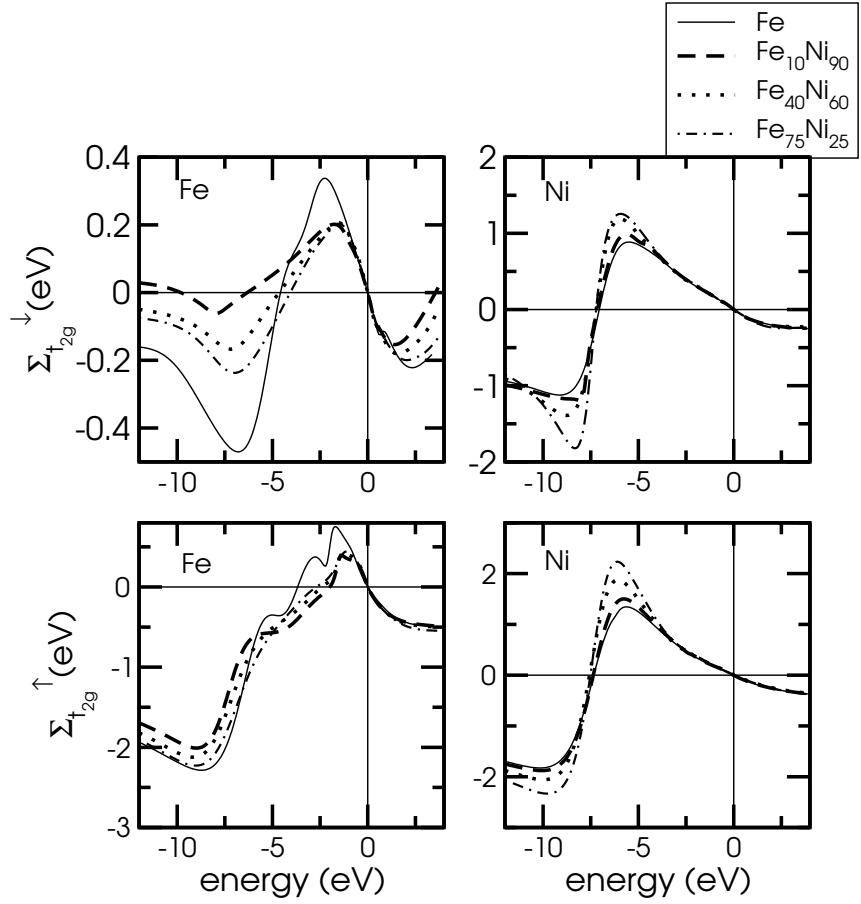}
\end{center}
\caption{Left: Concentration dependence of the real part of the spin
  resolved self-energy for Fe. Only results for $t_{2g}$ d-orbitals are shown. 
Right: Same as in the left panel but for Ni}
\label{FIG:FeNiSE}
\end{figure}

Finally, in Fig.~\ref{FIG:FeNiSE} the resulting concentration dependence of the
self-energy is shown. We present the results for the real part of the
self-energies for Fe and Ni atoms for the $t_{2g}$ symmetry (the
results for the $e_{g}$ symmetry are similar and hence not
plotted). It is interesting to note that the slope of the self-energy
near Fermi level $Z=\frac{d\Sigma}{dE}\vert_{E=E_F}$ which defines the
mass renormalisation and leads to the narrowing of the band
practically does not depend on the concentration. On the other hand,
for the high energy part of the self-energy one sees rather noticeable
differences giving raise to the changes in the sattelite structure.

\section{Summary}
A scheme has been presented that allows to combine the KKR band
structure method and the LDA+DMFT approach to deal with correlated
systems. Its applicability has been demonstrated by results for
ferromagnetic bcc-Fe and fcc-Ni. For this systems a good agreement
with previous LDA+DMFT methods has been found. In addition we combined
LDA+DMFT
scheme with the CPA to deal with disordered alloy. As an example we
presented results for fcc-Fe$_{x}$Ni$_{1-x}$ disordered alloy.

%\section*{Acknowledgement}

%
%%%%%%%%%%%%%%%%%%%%%%%%%%%%%%%%%%%%%%%%%%%%%%%%%%%%%%%%%%%%%%%%%%%%%%%%%%%
%                         REFERENCES
%%%%%%%%%%%%%%%%%%%%%%%%%%%%%%%%%%%%%%%%%%%%%%%%%%%%%%%%%%%%%%%%%%%%%%%%%%%
%
%\bibliographystyle{prsty}
%\bibliography{/opt/ak/bib/akhelit,lulu}

\newpage

\newpage
\newpage
\begin{figure}
\begin{center}
  \includegraphics[height=10cm]{fig2.eps}
\end{center}
\caption{Spin resolved density of states of fcc-Ni as calculated in
  the LDA (dashed line) and LDA+DMFT (full line) using the
  KKR-method. (DMFT parameters: U=3eV, J=0.9eV T=400K) }
\label{FIG:NiDOS}
\end{figure}
\newpage
%FFFFFFFFFFFFFFFFFFFFFFFFFFFFFFFFFFFFFFFFFFFFFFFFFFFFFFFFFFFFFFFFFF

\newpage

\begin{table}
\caption{
The DMFT parameters average Coulomb interaction U, exchange
energy J and temperature T used in the calculations for bcc-Fe,
fcc-Ni and fcc-Fe$_{50}$Ni$_{50}$. In addition the theoretical
spin magnetic moments as calculated by the LDA and the LDA+DMFT methods
are shown for bcc-Fe and fcc-Ni. Magnetic moments for
fcc-Fe$_{x}$Ni$_{1-x}$ alloy are presented in Fig.~\ref{FIG:FeNiMAG} %
}\label{TAB1}
\begin{tabular}{r|c|c|c|c|c}
 &U(eV)&J(eV)&T(K)&$\mu_{spin}^{LDA}(\mu_{B))}$&$\mu_{spin}^{DMFT}(\mu_{B))}$\\ 
\colrule
bcc-Fe&2.0&0.9&400&2.29&2.28\\
fcc-Ni&3.0&0.9&400&0.59&0.57\\
Fe in Fe$_{x}$Ni$_{1-x}$&2.0&0.9&400&&\\
Ni in Fe$_{x}$Ni$_{1-x}$&3.0&0.9&400&&
\end{tabular}
\end{table}

\end{document}